# 1953: An unrecognized summit in human genetic linkage analysis*

E.A. Thompson†

*Department of Statistics, University of Washington,
Box 354322, Seattle, WA 98195, U.S.A.
e-mail:* `eathomp@u.washington.edu`
*url:* `http:www.stat.washington.edu/thompson`

**Abstract:** This paper summarizes and discusses the methodological research in human genetic linkage analysis, leading up to and following from the paper of C. A. B. Smith presented as a Royal Statistical Society discussion paper in 1953. This paper was given as the Fisher XXVII Memorial Lecture, in Cambridge, December 4, 2006.

## 1. Introduction

### 1.1. Background

First a word of explanation about the title. The subtext reason for this particular title is of course that 1953 was the year Everest was climbed. That is, I believe, the first non-family event that I can recall, but it is far from an unrecognized event. In the world of science, it was the year of Watson & Crick (43) and the start of the DNA era, also far from unrecognized, and Dr. Edwards tells me that 1953 was also the year in which R.A. Fisher was knighted, so also an important year for him. However, the reason for my emphasis on 1953 today is that that is the year of C.A.B.Smith's discussion paper on human genetic linkage detection in JRSS(B) (33).

We will come back to that in due course, but first, it is of course a great honour to give a Fisher Memorial Lecture, and also an opportunity to repeat one's favourite Fisher quotes. Here are two of mine, both relating to the teaching of Statistical Scientists. The first was written in 1933 shortly after he became Galton Professor at University College London, to J.B.S. Haldane who was then Professor of Genetics.

---

*This paper was accepted by Elja Arjas, the Executive Editor for the Bernoulli Society.
†Supported in part by NIH grant GM-46255.

> Supposing mathematically trained students come to me ...., knowing nothing and not very willing to know anything of experimentation with living material, can I make them attend lectures in your department .... as one step towards apprehending the kinds of reasoning used by experimenters.

This one resonates with me, after 7 years of building a program in Statistical Genetics, even at University of Washington, which is one of the best and most supportive places for such an endeavor. Students can sometimes be their own worst enemies.

The second quotation is from a letter to John Wishart, dated October 27, 1949, agreeing to serve on a Faculty Board Committee to review the proposal for the establishment of the Cambridge Diploma in Mathematical Statistics:

> There is no wide or urgent demand for people who will define methods of proof in set theory in the name of improving mathematical statistics. There is a widespread and urgent demand for mathematicians who understand that branch of mathematics known as theoretical statistics, but who are capable also of recognising situations in the real world to which such mathematics is applicable.

That is as true today as it was then, and it is a particular pleasure to be giving this talk here at the Isaac Newton Institute, where real-world mathematical and statistical science are a priority.

In the event, I believe the Cambridge Diploma has educated many Statistical Scientists of the kind Fisher would have approved of, including, I hope, myself. My own links to Fisher are indirect, but the first was as a diploma student when I did my applied project under the supervision of Dr. Anthony Edwards, then finalizing the first edition of his book "Likelihood". I then became Dr. Edwards' PhD student, and Dr. Edwards was the last Cambridge undergraduate admitted to Part II Genetics under R. A. Fisher. After that, I did one year of postdoctoral research with Luca Luigi Cavalli-Sforza, who had been a research associate in Cambridge with Fisher around 1950. In fact, the first Fisher Memorial Lecture I attended was the VIth, in London in June 1974, at which Professor Cavalli-Sforza spoke on cultural versus biological evolution, an area for which he has become famous (7). I have to admit, however, that the main reason I remember that event is that it was the day after I submitted my PhD thesis.

Nonetheless, this is, in some sense, my third Fisher lecture. Fisher was a founding member of the Biometric Society, now International Biometric Society (IBS), and 1990 was the centenary of R. A. Fisher's birth. IBS had a memorable meeting in Hungary that year, starting the weekend that the currencies of the former East and West Germanies German currencies were unified, and delegates from Denmark and northern Germany could first come by train directly through Berlin to Budapest. The meeting was also memorable for the Fisher Centenary talks, at which I gave one in the opening plenary session on *Fisher's contributions to genetical statistics* – a rather large topic to which one lecture cannot do justice (40). Then, in 1994, at the Joint Statistical Meetings in the USA, I gave my second, and, I believe, much more successful lecture on *Likelihood and Linkage* (41). As a result, I cannot add anything new to that topic in the context only of R. A. Fisher. That then is why, for this talk, also about likelihood and linkage, I would like to focus on the contributions of C. A. B. Smith.

### *1.2. Cedric Austen Bardell Smith: 1917 – 2002*

Unlike Fisher's biography, Cedric Smith's is short, despite his long life. He started at Cambridge, for which he held a life-long affection, first as an undergraduate (1935-38) at Trinity College, reading the Mathematics Tripos, then as a Ph.D.Student (1939-42) in the early days of the Statistical Laboratory. There he was supervised by Bartlett, Wishart, and Irwin. Then, until the end of the war, he worked as a porter at (the Old) Addenbrooke's Hospital. In 1946, he joined the Galton Laboratory, at University College London, where Fisher had been the Galton Professor until 1943. Fisher was by then returning to Cambridge, but Haldane was still Professor of Genetics. There at University College, C.A.B. Smith remained for his long and active academic life, becoming Weldon Professor of Biometry in 1964.

Anthony Edwards wrote the obituary of C.A.B Smith for the Royal Statistical Society, and made two comments that resonated with me

> .. logical precision and intellectual honesty ..
> .. lasting influence more on the way people think than on technical details...

I hope to demonstrate those two characteristics of Cedric's work today.

Although I did not know Fisher, I did know C.A.B.Smith for over 25 years, as a mentor and friend. I met him first while I was a graduate student here in Cambridge. He was, in fact, the external examiner for my PhD thesis. More importantly, both for me and for many others, he was co-Editor of Annals of Human Genetics for many years. Four of my first six papers were published there, and he reviewed those papers with meticulous but benign care: no logical detail escaped him, and the papers are much the better for it. Cedric nominated me for the ISI in 1981, and he was one of my references in the mid 1980s when I decided to leave Cambridge, although I think he could not understand why I would wish to do so.

However, I remember Cedric best from those years as a regular and active participant at the annual Mathematical Genetics meetings. I remember the one in Liverpool in 1976, to which I will return later. I also remember particularly his presence at the two which I organized in Cambridge in 1978 and in 1985, and his evident enjoyment in visiting Cambridge. The 1985 meeting was not only the last I attended until the recent 30-year anniversary of their beginning, but also occurred just as I was about to interview at University of Washington (UW), for which Cedric had been one of my references. In one of the longer discussions I had with him, I recall his prediction that I would not return to UK academia, and of course he was right. Although I cannot match Cedric's long tenure at University College, my 21 years at UW seem to have gone very quickly.

## 2. The development of human genetic linkage analysis

### 2.1. Base-camp: The foundations.

Now, to reach the summit in 1953, we must first return to the base camp of human genetic linkage analysis, which is undoubtedly the work of R.A.Fisher, J.B.S.Haldane and L.S.Penrose in 1934-35. These three scientists realized that the methods of gene mapping already well established for experimental organisms could be applied to human genetic data, but also that there were difficulties specific to the fact that human data are of necessity observational. All three focused on linkage detection, in effect testing a null hypothesis of absence of linkage, although the term *null hypothesis* was not yet in use. Fisher (14) described linkage detection as *"...evidently that which will first require discussion as data .. become available"* As we shall see later, absence of linkage has an almost unique status as a null hypothesis; it has been referred to by Professor John H. Edwards as "the only true null hypothesis in biology" (pers.comm.).

The contribution of Penrose (32) is a little different, and in some ways simpler. His test statistic is based on correlations between traits in similarity of phenotypes of relatives. If there is linkage, sibs similar for one trait will be similar for the other. Haldane (18) used the method-of-moments u-statistic estimating equations of Bernstein (3) to form a test statistic for proportions of observations being in accord with those expected in the absence of linkage. Thus he too focussed on linkage detection, but pointed out that the same equations may be used for estimation of recombination fractions. However, this approach provides efficient estimates only close to the null hypothesis, where estimation is usually of little interest. Fisher (14) noted the estimation potential of Haldane's approach, but also focussed on detection. He also started from Bernstein's estimating equation, but related it to the score and information of his method of maximum likelihood (13). That is, to standardize his test statistic, he derived and used a variance appropriate for any linkage value, not just the null hypothesis. In turn this leads to results for what we would now call the power to detect linkage as a function of the recombination fraction.

### 2.2. Camp 2: Likelihood-based scores for linkage detection

The next major advance came in 1940, when Fisher (16) developed his approach making clearer the relationship to efficient scores in likelihood-based testing procedures. This was taken up by D. J. Finney in a series of papers ((12), and following). Finney (12) considered general multinomial data, with cell probabilities $p_i(\eta)$, $i = 1, ..., k$, and a null hypothesis $H_0 : \eta = 0$. The log-likelihood contribution for each observation in category $i$ is $\log p_i(\eta)$, the score contribution is

$$\frac{\partial}{\partial \eta} \log p_i(\eta)|_{\eta=0} \ = \ p'_i(0)/p_i(0).$$

Finney (12) expands the likelihood contribution $p_i(\eta)$ about $\eta = 0$ to obtain

$$L_i = p_i(0) (1 + \lambda_i \eta + O(\eta^2))$$

Efficient scores sum over independent observations, so that at $\eta = 0$ the total score becomes $= \sum_{\text{obs}} \lambda_i$, and the information for $\eta$ is $\sum_{\text{obs}} p_i \lambda_i^2$.

This series of papers is a major excursion up the mountain, but it has its limitations. First it is likelihood, not log-likelihood as would be so for another 10 years. The other is a problem that all the early statistical geneticists struggled with: the fact that, in genetics, not only do we have dependent observations, but that it is the dependence that is all-important. The unit of observation is the pedigree, or in their case the nuclear family. The number of categories $k$ of the multinomial is the number of different joint trait and marker data configurations on each pedigree structure. The likelihood must be computed, and in the days before computers the scores must be tabulated, for each family size and data configuration, and for each trait model.

### 2.3. Camp 3: Linkage likelihoods for estimation

The next large step came with the work of Bell & Haldane (2) and Haldane & Smith (20). In some sense this is a digression from the main theme, as these papers address estimation of recombination fractions between X-linked loci. However, in supporting the advance to the 1953 summit this work is important for three reasons. First, Bell & Haldane (2) introduced use of likelihood to the estimation of recombination fractions from data on human pedigrees of, in principle, any structure. Likelihoods are multiplied over disjoint pedigree structures, and the overall MLE of recombination fraction estimated from the overall likelihood function. Second, Haldane & Smith (20) went further, using the likelihood ratio $L(\chi)/L(1/2)$ at recombination fraction $\chi$ compared to its value under no linkage ($\chi = 1/2$). Further, the issue of the fact that there are substantial prior odds on absence of linkage is now explicitly addressed: the prior for the recombination fraction includes a point mass at $\chi = 1/2$.

In terms of linkage analyses using pedigree data, however, the most important advance of Haldane & Smith (20) is the way they compute their probabilities of observed data on four- and five-generation pedigrees. They explicitly condition on phased genotypes of pedigree members who divide the pedigree into two parts, noting the consequent conditional independence of data on the two parts of the pedigree. Hence, they accumulate the probability sequentially over the pedigree. The argument is identical to that used by Elston & Stewart (11) in the now-famous Elston-Stewart algorithm. It is the basis not only of the computational algorithms that came to be known in the human statistical genetics world as "(pedigree) peeling", but also of computations used in statistical analyses of graphical models (28).

This brings me back to the Mathematical Genetics meeting in Liverpool in 1976. There, Professors Chris Cannings and C. A. B. Smith independently gave the first talks on pedigree peeling computations for general pedigrees (29), although I suspect few of the audience realized they were presenting equivalent algorithms. These algorithms were generalized further by Cannings et al. (5; 6), and helped to bring statistical genetics back into the statistical mainstream when the same ideas were applied to general graphical models by Lauritzen & Spiegelhalter (28).

### 2.4. Camp 4: Lod scores and inference

We come now to the final supporting camp before the summit: the use of lod scores for inference (1). I cannot here do justice to Barnard's JRSS(B) Discussion Paper, but mention here only that this is the paper where the odds-ratio and lods were first introduced as general tools for inference. Barnard's b-lods (or backwards lods) are what we now know as the log likelihood-ratio. As a JRSS discussion paper, there were of course many discussion comments but of most interest to us here are those of C. A. B. Smith. Clearly, he was intrigued and excited by these ideas. He makes three main comments, which indicate he may have already been thinking in terms of application of these ideas to the detection of genetic linkage from pedigree data. (1) He likes the log-odds (lods) in preference to the odds-ratio, citing the additivity of lods over independent data structures. (2) He compares favourably the approach of Barnard (1) with that of the Neyman-Pearson theory, noting that while that framework has a rigid acceptance/rejection framework, *"lods ... has not this rigidity ... yet gives ...control of error."* (3) He raises again the problem of the composite alternative hypothesis, as well as the fact that weight should maybe be assigned to the null hypothesis, suggesting a prior distribution and a likelihood function integrated with respect to this prior.

## 3. Lod scores for linkage

### 3.1. The 1953 summit

So now we come to Smith's 1953 paper (33) also a JRSS (B) discussion paper. We have the four supporting elements that lead to linkage lod scores: (1) the foundation work of Fisher (14) and Haldane (18); (2) scores and information for testing for linkage related to derivatives of the likelihood (16; 12); (3) likelihoods computed on pedigree structures (20); and (4) lod scores as a general tool for inference (1).

In order not to interrupt the flow, I am going to go straight to the lod score arguments of Smith (33), but I shall later come back to many other key points made in the paper. First, as did others, he reparametrizes, using $\eta = 1 - 2\chi$, so that the null hypothesis is $H_0: \eta = 0$ and $0 \leq \eta \leq 1$. As did Finney (12) he considers multinomial data in categories $i$ and expands

$$p_i(\eta) \;=\; p_i(0)(1 + a_i\eta + O(\eta^2)),$$

noting again $p'_i(0) = a_i p_i(0)$ giving Fisher's score $a_i$ for an observation in category $i$. Then however, he converts to log-likelihoods, giving the contribution of each $i$-type observation to lod score as

$$\ell_i = \log_e p_i(\eta) - \log_e p_i(0) \approx a_i \eta.$$

Here the equation is given only to first order, although Smith (33) is much more careful about the higher-order terms. The lod score for small $\eta$ is then lod $= \sum_i x_i a_i \eta$ for $x_i$ $i$-type observations, and rejecting $H_0$ if lod score large is large is equivalent to Fisher's efficient score $\lambda = \sum_i a_i x_i$ being large.

Although Smith (33) shows this first-order equivalence with Fisher's scores, he also makes clear that he is following Barnard (1) in viewing the lod score as a tool in its own right, commenting that one may use *"natural or common lods according to convenience."* (In the days of log tables, before computers, common base-10 logarithms were indeed more convenient for most purposes.) Rather than tabulate scores, the goal is now to compute log-likelihoods on pedigrees.

As a JRSS(B) Discussion paper, there were many comments. Those of J.B.S.Haldane are perceptive. He said *"... for many purposes the lods approach may be more fruitful"* and *"...[this is a] novel approach which is going to be absolutely fundamental."*. The subsequent 50 years have proved him right.

### 3.2. Power and type-1 error

The criterion of a (base-10) lod score of 3 for detection of linkage was introduced by Morton (30). It has proven an immensely successful criterion, providing almost no false positive results for simple genetic traits, and leading eventually to the the construction of genome-wide human genetic maps (27). However, it is neither a necessary consequence of the sequential probability ratio test (SPRT) approach of (30), nor directly chosen to ensure a small type-1 error rate. Indeed, there is no fundamental difference between a sequential and non-sequential approach to linkage detection once the lod score is chosen as the test criterion. In his introduction, Morton (30) states

> ... current methods to detect human linkage.... u-scores, Penrose sib-pair method, the [likelihood] ratio method... Smith (1953) has shown that they are all really different forms of the nonsequential probability ratio test.

while Smith (35) in an uncharacteristically sharp comment says

> It seems to me that the use of a ... sequential stop-rule ... is not appropriate in linkage work and confers no advantage: on the other hand Bayes' theorem can be quite easily applied ... and gives a more satisfactory answer from both the theoretical and practical point of view.

The criterion of a lod score of 3 for detection of linkage relates rather to the probability a detected linkage is true; in modern terminology the false discovery rate (FDR). A prior probability of linkage is therefore required, but this need not be considered a Bayesian procedure. Rather the prior probability relates to the process of sampling the loci that are to be tested for linkage. In Smith (33) the argument implicitly assumes a 5% type-1 error rate, and explicitly assumes sufficient information that all true linkages will be detected. He suggests that, as an approximation, only 1 in 24 of random pairs of loci will be linked, so that in every 100 pairs tested there will be 4 true linkages and 5 false ones, resulting in *"more false positives than true ones."* Morton's argument (30) is more precise. He assumes the prior probability of linkage is $\pi \approx 1/20$, and derives

$$\text{FDR} \;=\; \frac{\alpha(1-\pi)}{(\alpha(1-\pi)+\pi W)} \;=\; \frac{19\alpha}{19\alpha + W}$$

where $W$ is the average power.

Morton (30) used the standard SPRT formulae relating type-1 error probability $\alpha$ and power $(1-\beta)$ to the lod-score bounds for acceptance and rejection of linkage. However, the basis of these formulae is not restricted to the sequential approach. It is here of interest to note the contribution of G. Barnard's to the Discussion of Smith's 1953 paper (33). As often, the JRSS Discussion comments contain as much of interest as the main papers, and Barnard gives an improvement to the error bound given by Smith, saying:

> [It is a] general result that ... odds of error ... equal to the conditional mean value in the critical region of the likelihood ratio. [This is] true regardless of whether or not sampling has been sequential, or ..

To clarify this comment, let $f_0()$ and $f_1()$ be the two probability densities for $x$ under the null and alternative hypotheses. Then the type-1 error and type-2 error probabilities are

$$\alpha \;=\; \int_C f_0(x)dx, \quad \text{and } 1-\beta \;=\; \int_C f_1(x)dx,$$

where $C$ is the critical region for the test. Barnard refers to the ratio $\alpha/(1-\beta)$ as the odds of error, and

$$\begin{aligned}
\alpha/(1-\beta) &= \int_C f_0(x) / \int_C f_1(x)dx \\
&= \int_C (f_0(x)/f_1(x))f_1(x)dx / \int_C f_1(x)dx \\
&= \text{E}_1(f_0(X)/f_1(X) \mid X \in C)
\end{aligned}$$

Since the null hypothesis is here in the numerator, we will reject $H_0$ if the likelihood ratio is small, and the critical region $C$ will be of the form $f_0/f_1 \leq 1/A$, and hence $\alpha/(1-\beta) \leq 1/A$. For the SPRT, we "ignore overshoot" and set equality in the inequality, obtaining $\alpha/(1-\beta) = 1/A$ or $A = (1-\beta)/\alpha$, and equivalently $B = \beta/(1-\alpha)$.

The main point of difference between these 1950s approaches, and what all these authors are struggling with is the issue of the *alternative hypothesis*. Whereas the Neyman-Pearson approach adopted a fixed alternative for which power is maximized subject to a given type-1 error, Fisher preferred not to consider the alternative hypothesis, testing only goodness of fit to a null hypothesis. Haldane & Smith (20; 33) balanced between the two extremes, wanting to weight values in the alternative to compute an integrated likelihood or posterior probability of linkage. Morton (30) has the interesting proposal of choosing the specific alternative to form the SPRT according to the amount of data available, in such a way as to keep approximately constant the power at that alternative. Thus, if very large amounts of data are available, an alternative very close to the null is considered; for less data, the alternative hypothesis is further from the null hypothesis. In the context of genetic linkage detection, these arguments are nowadays moot. If there are genes affecting trait values, there will be tight linkage to some marker in a genome scan, and the issue is one of multiple dependent tests rather than of the composite alternative hypothesis.

## 4. Further aspects of human genetic linkage analysis

### *4.1. Smith (1953): the background*

We return now to the early part of Smith's paper (33), which has, in many ways, a very modern flavor. He discusses effects on phenotype of environment, genetics and of gene-environment interaction. For complex traits, he discusses the issues of delayed age-of-onset, penetrance, and problems of variation in severity and diagnostic criteria. On the genetic side, he notes that there was at that time little direct experimental evidence of chromosomal inheritance in humans. His arguments are made by analogy with plant and animals genetic mapping approaches, and he comments, as had others before him (14; 18) on the problem of using observational data as opposed to data from experimental organisms. He discusses genetic complexities, such as inversions, translocations, and deletions, and notes the possibility of using inherited deletions as genetic markers. Finally, he addresses the issue of genetic heterogeneity. Whereas Haldane & Smith (20) had addressed allelic heterogeneity, which poses little problem for linkage analysis other than in the estimation of disease allele frequencies, (33) addresses locus heterogeneity, although also to some extent dismissing it since

> ... the significance test gives linkage with any one of these [loci].

Next, Smith (33) addresses the issue of associations, both at the population level and among relatives. In the context of linkage detection, he discusses when and where there is information for inference, noting again that the unit of observation is the whole pedigree. In the population at large, he states that after a few generations linkage disequilibrium (LD) will be slight, using this explicitly as the justification for assuming equal probability for phase in founder members of a pedigree. However, he also notes that if linkage is very tight, LD will be maintained over many generations and can provide evidence for the detection of linkage. While Fisher (17) had used the same argument of population haplotype frequencies to order the three rhesus loci, this seems to be the first suggestion of the use of LD for linkage detection, although of course at that time today's dense genomewide maps that make this a practical proposition were not available. In the same vein, (33) addresses also family-based associations. He notes that if and only if there is LD in the population a single parent-offspring pair can provide evidence of linkage, foreshadowing the transmission disequilibrium test (TDT) (38), but that otherwise two offspring are required.

Smith (33) then moves to *ibd*-based methods for linkage detection, summarizing the sib-pair method of Penrose (32), noting that this approach does not require specification of a trait model. Finally, he moved to inbreeding. Haldane (19) had noted that variation in the level of inbreeding within a population causes associations between recessive characters at the population level, and that these associations are increased by linkage between the loci underlying such recessive traits. Whereas Haldane (19) had been primarily interested in the evolutionary aspects of these associations, Smith (33) demonstrated how such associations could be used for linkage detection. He thus predicted the method of *Homozygosity Mapping* (25), used successfully in the 1990s to map many rare recessive traits, although, as with the TDT, the genome wide maps that would make this approach practical would not be available for 30 years.

A modern review paper (26) describes four methods for mapping the genes for complex traits, summarizing then as: (1) Lod scores for linkage; (2) Associations in populations; (3) *ibd*-based methods; and (4) mapping of quantitative trait loci in experimental organisms. This is a useful review paper, but there is little integration of the four approaches, and some of the papers cited remind me of another R. A. Fisher quote (15)

> .... it is usually understood that the conclusions drawn from experimental results must rest on a detailed knowledge of the experimental procedure actually employed. Nevertheless, in human genetics especially, statistical methods are sometimes put forward, and their respective claims advocated with entire disregard for the conditions of ascertainment.

In comparison, Smith (33) covers, at some level and for discrete Mendelian traits, all four of these areas, starting with mapping in experimental organisms, proceeding to discuss population associations, then developing lod scores, and finally coming back to *ibd*-based methods in the form of sib-pair analyses and homozygosity mapping. More importantly, he presents them not as four distinct areas, but as aspects of the single scientific problem of linkage detection, with a single overall (likelihood) approach to solution. To quote Morton (30) again

> ... current methods to detect human linkage.... u-scores, Penrose sib-pair method, [likelihood] ratio method... Smith (1953) has shown that they are all really different forms of the nonsequential [likelihood] ratio test.

and, had he been writing 40 years later, he could have added homozygosity mapping (25), TDT (38) and other family-based association tests that may be the best hope for mapping the genes associated with complex traits. Of course, much has been achieved since 1953, but just as the seeds of Smith (33) are in the work of Fisher (14; 17) and Haldane (18; 19), so also are the seeds of modern linkage methods in Smith (33).

### 4.2. Later contributions of C. A. B. Smith

Smith (33) is a summit, and, together with Morton (30), set the stage for much subsequent work in linkage analysis for the following 40 years. However, it is far from the only contribution of C. A. B. Smith, and there are two for which, perhaps, he is better known.

The first is the use of "Gene counting" for the maximum likelihood estimation of population allele frequencies. The method was introduced in a paper with colleagues (8), but less well known is Smith (34) in which he studied convergence, relating it to conditional variance of the latent alleles counts given the data. Of course, "Gene counting" is a special case of the EM algorithm (9) and EM algorithms are widespread in Statistical Genetics, but many of the results of Dempster, Laird and Rubin (9) are first given for the gene-counting case by Smith (34).

Second, and to those working in human genetic linkage analysis perhaps best known (31) is Smith's test for locus heterogeneity (36). First mentioning locus heterogeneity as "not a problem" for linkage detection (33), he later introduced his test (36) which is still widely used.

Last but not least, is his work on algorithms for the computation of probabilities on pedigree structures, the conditional independence logic being already present in Haldane & Smith (20) and made explicit in Smith (37). Although presented in the context of a Mendelian model and a specific example, Smith (37) is more general than Elston & Stewart (11) in that computations may be made both upwards and downwards over the pedigree structure. The same logic was used by Cannings et al. (5; 6) to extend to pedigrees of arbitrary structure and more complex genetic models.

### *4.3. Elods: the expected lod score*

It is perhaps surprising that the *elod*, or expected lod score, finds no place in Smith (33) nor in his later work. The Kullback-Leibler information

$$KL(f_1; f_0) \;\; = \;\; \mathrm{E}_0(\log(f_0(X)/f_1(X)))$$

had already been introduced as a measure of the information to distinguish probability densities $f_0$ and $f_1$ when $f_0$ is true (24). However, the relationship to Fisher information and likelihood seems not to have been appreciated until the work of Kempthorne (22) and Kendall (23).

Whereas for most applications of likelihood ideas and inference tools, statistical genetics, and indeed linkage analysis, have been at the forefront, *elod*s seem not to have entered the Statistical Genetics literature until Thompson (39) and not in relation to the detection of genetic linkage until Thompson et al. (42). This is not because they are not applicable in this context; they have become an important tool for assessing information for linkage detection (31). There are two likely reasons for the relatively late appearance of *elod*s in statistical genetics. The first is computational. To compute a lod score on a pedigree it is necessary to sum over all possible genotypic configurations on a pedigree that give rise to the observed data. To compute an *elod*, it is necessary to also sum over all possible data configurations. Even in 1978 this was a daunting task even on small pedigrees: the *elod*s of (42) were estimated by Monte Carlo, since exhaustive ennumeration was considered impractical. Another reason may be the statistical inference framework adopted by Smith (33). The *elod* requires a specific alternative hypothesis $f_1$. We have already seen how the earlier researchers struggled with the problem of the alternative hypothesis, and in particular how C. A. B. Smith in his discussion of Barnard (1) expressed reservations about "rigidity" of the Neymann-Pearson framework with its specific alternative.

## 5. Summits in Biology and Technology

New ideas in basic inference for Statistical Genetics (as opposed to Population Genetics) since 1953 have been few and far between. However, there have, of course, been advances in methods, keeping pace with the huge advances in Biology and Technology which have required new computational tools and algorithms.

When C.A.B.Smith wrote his 1953 paper, it was still believed that the human cell nucleus had 48 chromosomes, and he repeatedly said that the direct evidence for genes/chromosomes etc. in humans was slight, and that he argued by analogy with much-better-understood plant and animal genetics. Since then we have had decades of revolutionary change:

- *1955-1965:* Very soon after, indeed starting in the same year (43), we had a decade of change in understanding of DNA transcription and translation, establishing the central dogma of DNA sequence to RNA to Protein.
- *1965-1975:* Then came the revolution in computing technology, making possible computations of probabilities and likelihoods on complex data structures.
- *1975-1985:* Then came DNA markers, from the RFLPs of (4) to the SNPs of today, providing genetic marker maps of the human genome, and making possible the idea of a genome-wide scan for linkage detection.
- *1985-1995:* Next came the biotechnology revolution, with automated methods for DNA sequencing and marker typing, providing huge increases in data and potentially the power to address complex traits.
- *1995-2005:* Finally, has come the bioinformatics revolution, providing the human genome sequence, the HapMap, and gene expression data.

Each stage has raised new challenges for statistical genetics, and perhaps I should end with quote from another (often unrecognized) statistical scientist: Florence Nightingale. I started with two quote of R. A. Fisher concerning the teaching of statistics at University College, and in Cambridge. This one is was written when R. A. Fisher was less than one year old, and concerns the teaching of statistics in the University of Oxford. On 3 January 1891, Florence Nightingale wrote to her friend Benjamin Jowettt about the possibility of setting up a Chair of Statistics at Oxford (21):

> ... the enormous amount of Statistics [i.e. data] at this moment at their disposal is absolutely useless. Why? Because... [they] have received no education whatever on the point on which all ... must ultimately be based. We do not want a *neat arithmetical sum*. We want to know *what we are doing*. What we want first is not ... an accumulation of facts, but to teach [them] the *uses* of facts.

I believe this is as much a danger with genetic and genomic data now as it was for Public Health Data 110 years ago, and I also believe it is a sentiment with which both C.A.B Smith, and R. A. Fisher would have heartily approved. To return again to Dr. Edwards' comment in his RSS obituary of C. A. B. Smith:

> .... we want to know [understand] what we are doing ....
> .... a lasting influence on the way people think.

**Acknowledgment**

This R. A. Fisher Memorial Lecture was given while visiting the Isaac Newton Institute, as a Rothschild Visiting Professor of the University of Cambridge. I am grateful to the Issac Newton Institute for their hospitality, November-December 2006.

I am also grateful to Dr.A.W.F.Edwards for helpful discussions relating to the early history of linkage analysis (10), and for the information provided by his obituary of C. A. B. Smith, written for the Royal Statistical Society.